\documentstyle[twocolumn,seceq,epsfig]{jpsj}

\def\runtitle{double layers of CeSb}
\def\runauthor{Fumihiko {\sc Ishiyama} and Osamu {\sc Sakai}}

\title
{
Theory on 
the Stability of the Ferromagnetic Double Layer Structure 
and on the Peak Structure of the Magneto-Optical Spectra
of CeSb
}

\author
{ 
Fumihiko {\sc Ishiyama}%
\footnote{Contact address: NTT Energy and Environment Systems Laboratories, Tokyo 180-8585}
and Osamu {\sc Sakai}$^{1,}$
}

\inst
{
Department of Physics, Tohoku University, Sendai 980-8578 \\
$^1$Department of Physics, Tokyo Metropolitan University, Hachioji 192-0397
}

\recdate
{
--------------
}

\abst
{
We propose the $pf$+$pd$ mixing model for CeSb
to explain the stability of the  ferromagnetic double layer structure 
in the magnetic ordering.
The $pd$ mixing causes the saddle type singular points, neighboring the $\Delta$ axis,
for the bands which gain energy through the $pf$ hybridization with the 
occupied $f$ state. 
The peak of the density of states due to this combined effect of 
the $pf$ mixing and the $pd$ mixing enhances the stability of the double layer structure.
The same combined effect also causes the saddle type singular points 
in the joint density of states of the optical transition.
The peak structure of the magneto-optical spectra 
which has been observed in experiments is explained 
by the present model.
}

\kword
{
CeSb, magnetic order, double layer structure, optical conductivity, $pf$ mixing model
}

\begin{document}
\sloppy
\maketitle

\section{Introduction}

Cerium monopnictides (Ce$X_{\rm p}$, $X_{\rm p}$=P, As, Sb, Bi) are 
semimetals with the NaCl type lattice.
Their electronic bands have hole surfaces originated from atomic $X_{\rm p}$ $p$ states 
around $\Gamma$, and electron surfaces originated from Ce $5d$ states 
around three equivalent $X$\cite{Hasegawa85}. 
In the temperature ({\it T}) 
and the magnetic field ({\it H}) diagram\cite{KohgiA0,RossatMignod85},
Ce$X_{\rm p}$'s show various types of magnetic ordered states.
They consist of the  ferromagnetic layers
which stack antiferromagnetically.
Many types of stacking sequences appear in the phase diagram.
Among Ce$X_{\rm p}$'s, CeSb~\cite{RossatMignod85} 
and CeBi~\cite{Bartholin,RossatMignod83} have the type-IA 
antiferromagnetic ($++--$) phase with nearly full ($J_z \simeq \pm 5/2$) magnetic moment 
as the ground state of weak $H$ case.
Here `$+$' denotes the ferromagnetic layer with the magnetic moment parallel to the 
field, and `$-$' denotes the layer anti-parallel to the field.

The pair of ferromagnetic layers ($++$ or $--$) 
is widely observed as a unit of the magnetic structure in Ce$X_{\rm p}$'s.
We call it as the  `double layer structure', hereafter. 

The phase diagram of CeSb has been  qualitatively 
reproduced by the ANNNI model~\cite{BoehmBak,BakBoehm,Kasuyaetal}.
However, it is not clear how the parameters of the phenomenological model are
related to those of the microscopic model.

The origin of the ferromagnetic layer with nearly full moment was explained by
the $pf$ mixing model%
~\cite{KasuyaVI,Takegahara,Hiroko1,Hiroko2,Hiroko3,Hiroko4,Hiroko5,Hiroko6}
proposed by Kasuya {\it et al.}
The model also succeeded in explaining the origin of the antiferromagnetic stacking
of the ferromagnetic layers.
However, the stability of the double layer structure with sufficient energy gain
has not been obtained. 

The $pd$ mixing term between $X_{\rm p}$ $p$ bands and Ce $d$ bands  
is neglected in the energy calculation of the original $pf$ mixing model,
because the $pd$ mixing effect disappears on the $\Delta$ axis where the $pf$ mixing becomes maximum.
Later, it was pointed out that  the $pd$ mixing term plays important role\cite{Sakai85,Sakai87,Kaneta00} to 
explain the Fermi surfaces of CeSb in the ferromagnetic 
phase~\cite{Kitazawa88,Aoki85,Aoki91,Aoki95,Settai93,Settai94}
and $++-$ phase~\cite{Settai94,Settai96}.

In this paper, we employ  the $pf$ mixing model including the $pd$ mixing effect, 
and examine the stability of the double layer structure in CeSb.
Hereafter, we call the  model as the  $pf$+$pd$ mixing model.
It is shown that the $p$ bands, 
which are pushed up by the occupied $f$ state to gain the $pf$ mixing energy,
hybridize through the $pd$ mixing with the $d$ bands around the $\Delta$ axis.
This combined effect of the $pf$ mixing and the $pd$ mixing causes saddle type singular points 
neighboring the $\Delta$ axis, and substantiates the energy gain due to the $pf$ hybridization.
The double layer structure is stabilized by the combined effect.

We also study the magneto-optical spectra of CeSb.
Recently, the magneto-optical  spectra in various magnetic ordered 
states have been extensively studied experimentally by Pittini {\it et al.}~\cite{Pittini96a} and 
by Kimura {\it et al.}~\cite{Kimura00,Kimura00PB,KimuraPC}
The individual magnetic ordered state shows its own characteristics in the optical conductivity spectra.
The spectra will give us fine information on the electronic states in the magnetic 
ordered states and work as a good probe to study the characteristic band structure of the $pf$+$pd$ mixing model.
The peak structures observed in the experiments are explained by the saddle type singular points in the 
joint density of states, which are caused by the combined effect of the $pf$ mixing and the $pd$ mixing.

In the next section, we introduce the $pf$+$pd$ mixing model.
In section 3, the stability of the double layer structure is examined. 
In section 4, the optical conductivity in the magnetic ordered state is calculated.
Finally, we summarize the results obtained by the $pf$+$pd$ mixing model in section 5.

\section{Model}

Let us choose the $z$ axis along the magnetic polarization,
then the ferromagnetic layer is in the $xy$ plane.
We denote the layer which has the polarization parallel to the axis by the sign `$+$',
and the layer which has the polarization anti-parallel to the axis by the sign `$-$'.
The magnetic ordering is classified by the stacking sequence of layers along the axis.

We employ the following Hamiltonian,
\begin{eqnarray}
H &=&
H_f + H_p+H_d+H_{pd}+H_{pf}\label{Hamiltonian}
,
\\
H_f &=& 
 \sum_{{l}} \{ \sum_{t=\kappa, \nu}
 \varepsilon_f f^\dagger_t({\mib R}_{l}) f_t({\mib R}_{l})
\nonumber \\
&+& {U_{ff}} f^\dagger_\kappa({\mib R}_{l}) f_\kappa({\mib R}_{l}) f^\dagger_{\nu}({\mib R}_{l}) f_{\nu}({\mib R}_{l})
\}
,
\\
H_p &=& \sum_{{\mib k}} \sum_{t=\kappa,\nu}
 {\varepsilon_p({\mib k}) p^\dagger_t ({\mib k}) p_t({\mib k})}
,
\\
H_d &=&  \sum_{{\mib k}} \sum_{s=\alpha,\beta}
 {\varepsilon_d({\mib k}) d^\dagger_s ({\mib k}) d_s({\mib k})}
,
\\
H_{pd} &=&  \sum_{{\mib k}} \sum_{s=\alpha,\beta}  \sum_{t=\kappa,\nu}
  V_{pd}({\mib k}, s, t) p^\dagger_t ({\mib k}) d_s({\mib k})
\nonumber \\
  &+& h.c.
,
\\
H_{pf} &=&  \sum_{{\mib k}}
  \sum_{{l}}
  \sum_{t, t^\prime=\kappa,\nu}
  V_{pf}({\mib k}, t, t^\prime)
\nonumber \\
 &\times&
  e^{i {\mib k} \cdot {\mib R}_{l}}
  f^\dagger_t({\mib R}_{l}) p_{t^\prime}({\mib k})
\nonumber \\
  &+& h.c.
.
\end{eqnarray}
Here, $H_f$ stands for the $f$ states,
and ${\mib R}_{l}$ denotes site $l$.
The quantities $f_\kappa({\mib R}_{l})$ and $f_\nu({\mib R}_{l})$ are the annihilation operators 
of the $f$ electrons at ${\mib R}_{l}$ with $\kappa$ and $\nu$ components 
of the $\Gamma_8$ symmetry, respectively\cite{Hiroko3}.
These states have large magnetic moment along the $z$ axis,
and are considered as the main components of the ferromagnetic layers\cite{Kaneta00,Iwasa99}.
$\kappa$ and $\nu$ are ascribed to the $+$ layers and the $-$ layers, respectively.
$\varepsilon_f$ is the $f$ level, and $U_{ff}$ is the $f$-$f$ Coulomb potential.

$H_p$ stands for the $p$ bands which have hole surfaces around $\Gamma$.
$p_t({\mib k})$ denotes the annihilation operator of the $p$ band electron with symmetry $t$.
There are six sheets of $p$ bands.
Among them,
we consider only $\kappa$ and $\nu$ components of the $\Gamma_8$ symmetry
which mainly mix with $\kappa$ and $\nu$ components of the $f$ states.
The remaining two bands of the $\Gamma_8$ symmetry ($\lambda$ and $\mu$)
are treated as a reservoir, 
and the two bands of the $\Gamma_7$ symmetry are neglected 
because they do not have Fermi surface.

$H_d$ stands for the $d$ bands which have electron surfaces around $X$.
$d_s({\mib k})$ denotes the annihilation operator of the $d$ band electron
with the $xy$ symmetry and the spin index $s$.
The $xy$ bands mix with the $p$ bands with $\kappa$ and $\nu$ symmetry
in the energy region near the Fermi energy, $E_{\rm F}$.
We retain the $xy$ bands in $H_d$, and the other $d$ bands ($yz$ and $zx$) are
treated as a reservoir.
In the $d$ bands, we use the spin index $s$ ($\alpha$ and $\beta$) 
because the spin-orbit interaction is negligible.
$H_{pd}$ and $H_{pf}$ stand for the band mixing terms.

In the Hamiltonian, we have retained only the bands which are mainly affected by the
magnetic ordering characterized by the ferromagnetic layers in the $xy$ plane.
Therefore the cubic symmetry of the paramagnetic phase is lost at this stage.
The $\Delta$-symmetry axis along the $z$ axis is called as the $\Delta_Z$ axis, hereafter.

In the phase diagram, there appear layers which are denoted by `$0$'\cite{RossatMignod85}.
They have small sublattice magnetic moment and are considered as the layers
with the $\Gamma_7$ symmetry\cite{Iwasa99}. 
In this paper, we assume that the $\Gamma_7$ layer is inert for the band structure 
because the $pf$ mixing is weak for the $\Gamma_7$ states.

We introduce Schrieffer-Wolff transformation~\cite{Schrieffer} to avoid 
treating the $f$ states as the $f$ bands, 
and to improve the precision of the numerical process in the band energy calculation.
With this transformation for the ordered states,
the $pf$ mixing term is replaced as the effective $p$-$p$ term as follows:
\begin{eqnarray}\label{hpf-prime-general}
H_{pf}^\prime
&=&
-
\sum_{{\mib k}}
\sum_{t, t^\prime, t^{\prime\prime}=\kappa,\nu} 
~ \sum_{n=0}^{N-1}
\frac{1}{\varepsilon_f} 
\nonumber \\
&\times&
V_{pf}^{*}({\mib k}+n \Delta {\mib k}, t, t^\prime) 
V_{pf} ({\mib k}, t, t^{\prime\prime})
\nonumber \\
&\times&
\frac{1}{N} \sum_{l=0}^{N-1} n_{t}({l}) e^{-2 {\pi} i \frac{l n}{N}}
\nonumber \\
&\times&
p^\dagger_{t^\prime}({\mib k}+n \Delta {\mib k}) p_{t^{\prime\prime}}({\mib k}),
\end{eqnarray}
with $E_{\rm F}=0$
and 
$\Delta {\mib k} = (0,~0,~ {2 \pi}/({Na}))$.
$N$ is the period of the magnetic ordering, 
and
$\Delta {\mib k}$ is the basic  reciprocal lattice vector corresponding 
to the period of the magnetic ordering, where $2 a$ is the length of the edge of the cube of the fcc lattice.
$\Delta {\mib k}$ is obtained by choosing primitive translation vectors, 
one of which represents the magnetic ordering
and the other two vectors are $(2a, 0, 0)$ and $(a, a, 0)$ in the $xy$ plane of the fcc lattice.
Of course, the wave number vector ${\mib k}+n \Delta {\mib k}$ should be contained 
in the first Brillouin zone of the fcc lattice.
$n_{t}({l})$ is the occupation number of $t$ state on the $l$-th layer 
in the unit cell of the magnetic ordered state.
In eq.(\ref{hpf-prime-general}), the terms which are written as the single site energy
of the $f$ states are dropped.
They are common for $\kappa$ and $\nu$ (and also for the other states of $\Gamma_8$).
This cubic crystal field splitting term due to the $pf$ mixing should be included
when we compare the energy of 
the phases with various fraction of
the $\Gamma_8$ states.
Since the bases of the bands are restricted, only the terms which satisfy
the condition $t=t^\prime=t^{\prime\prime}$ 
in eq.(\ref{hpf-prime-general}) appear. 

The band parameters are fitted 
so as to reproduce the band structure of LaSb\cite{Hasegawa85}. 
\begin{table}[h]
\caption{Band parameters}\label{BandParameters}
\begin{tabular}{@{\hspace{\tabcolsep}\extracolsep{\fill}}c|rrrr}
(eV)       & ~$p$    & ~$d$     & ~$pd$    & ~$pf$    \\
\hline
$\varepsilon_a$        & -0.495  & ~2.526   & ~---     & ~---     \\
\hline
$\sigma_1$ & ~0.256  & -0.681   & ~0.766   & ~0.350   \\
$\pi_1$    & -0.020  & ~0.196   & -0.653   & -0.245   \\
$\delta_1$ & ~---    & ~0.000   & ~---     & ~---     \\
\hline
$\sigma_2$ & ~0.087  & ~0.000  & ~---      & ~---     \\
$\pi_2$    & ~0.104  & ~0.000  & ~---      & ~---     \\
$\delta_2$ & ~---    & ~0.023  & ~---      & ~---     \\
\end{tabular}
\end{table}
Table \ref{BandParameters} shows the band parameters used in the present calculation.
$\varepsilon_a$ is the atomic energy levels.
$\sigma_1$, $\pi_1$ and $\delta_1$ are the two center integral parameters for the nearest neighbor sites.
$\sigma_2$, $\pi_2$ and $\delta_2$ are the parameters for the next nearest neighbor sites.
The next nearest neighbor terms are included to reproduce the band dispersion curve 
precisely on the $\Delta_Z$ axis where the $pf$ mixing becomes maximum.
The parameter values of the $pd$ mixing term and the $pf$ mixing term are the ones 
used in the previous studies\cite{Hiroko2}.
In later calculation, we control the strength of the $pf$ mixing effect 
by changing  $-1/\varepsilon_f$.
Here, $\varepsilon_f$ is measured from the Fermi energy of the paramagnetic phase.

The energy of each ordered state is calculated from 
the density of states (DOS) by the equation
\begin{equation}
E=\int_{-\infty}^{E_{\rm F}} {\rm d}E^\prime 
(D_{\rm B}(E^\prime) + D_{\rm R}(E^\prime)) E^\prime.
\end{equation}
Here, $D_{\rm B}$ is the DOS of the bands which are treated explicitly in the Hamiltonian, 
and $D_{\rm R}$ is the DOS of the reservoir bands.
The reservoir is necessary to give the right Fermi energy $E_{\rm F}$ and the right carrier number.
As is shown later, $D_{\rm B}$ of the phases with double layer structure has a
peak splitting structure 
across the Fermi energy of the paramagnetic phase ($E_{\rm F} \simeq 0$ eV),
and this gives energy gain.
Without the reservoir, the Fermi energy rises up above 
the peak splitting structure, and the energy gain is not obtained. 
We calculate the DOSs ($D_{\rm B}$ and $D_{\rm R}$) 
by the tetrahedron method.
The calculated energy depends on the number of meshes in the ${\mib k}$-space $N_{\mib k}$.
We extrapolate the energy to the $N_{\mib k}\rightarrow \infty$ limit 
as a function of $1/N_{\mib k}$ from the calculated results of $N_{\mib k}=20, 30, 40, 50, 60$, 
and estimate the extrapolation error $\Delta E$ of the energy.

\section{Double Layer Structure}

In this section, 
we calculate the band energy of each magnetic ordered state, 
and show that the $pf$+$pd$ mixing model gives the stabilization of the double layer structure.

\begin{figure}
\epsfig{file=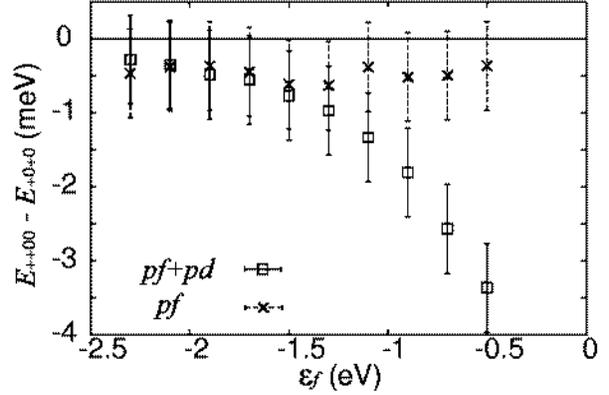,width=8cm}
\caption{
The energy difference between the $++00$ phase and the $+0+0$ phase $E_{++00}-E_{+0+0}$
as a function of the $f$ level $\varepsilon_f$. 
$\Box$ corresponds to the $pf$+$pd$ mixing model, 
and $\times$ corresponds to the $pf$ mixing model.
The error bars indicate twice of the extrapolation error $\Delta E$.
}
\label{f940}
\end{figure}

Figure \ref{f940} shows the energy difference between
the $++00$ phase and the $+0+0$ phase as a function of $\varepsilon_f$.
The energy of the $++00$ phase, which has the double layer structure, 
becomes lower than that of the $+0+0$ phase  in the $pf$+$pd$ mixing model
when the $f$ level $\varepsilon_f$ is shallower than $-1.5$ eV.
In contrast, the energy difference between the two is small in the $pf$ mixing model.
The magnitude of the energy gain which is obtained by the $pf$ mixing model 
is comparable to 
that of the previous work\cite{Hiroko5}.

We have also calculated the energy difference 
between the $++0000$ phase and the $+00+00$ phase based on the $pf$+$pd$ mixing model.
The former, which has the double layer structure, has lower energy when the $pf$ mixing is strong.
The $++000000$ phase has also lower energy than the $+000+000$ phase 
in the $pf$+$pd$ mixing model\cite{ORBITAL}.
The magnitude of the relative stability of the $++0000$ phase and the $++000000$ phase is about $1/3$
of that of the $++00$ phase in the range shown in Fig.\ref{f940}.
We use the value $\varepsilon_f=-0.7$ eV hereafter to analyze the origin of the 
stability of the double layer structure.

We note that the $t=\kappa$ component hybridizes only with the $s=\alpha$ (up spin) component
of the $d$ bands in the present band scheme, because the other components 
which can weakly hybridize are dropped in eq.(\ref{Hamiltonian}).
The $\kappa$ bands and the $\alpha$ bands do not hybridize 
with the $t=\nu$ and $s=\beta$ (down spin) components.
Therefore, we call the former part as the $+$-spin bands and the latter as the $-$-spin bands.
 
\begin{figure}[hbtp]
\epsfig{file=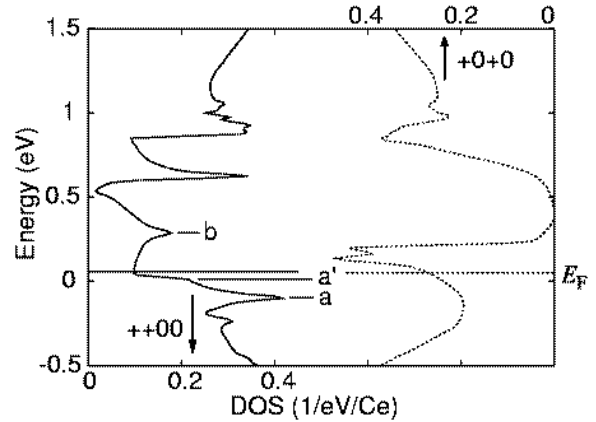,width=8cm}
\caption{
The DOS of the $+$-spin bands for the $++00$ phase (solid line) 
and that for the $+0+0$ phase (broken line).
Horizontal lines are $E_{\rm F}$ for the $++00$ phase (solid line) 
and that for the $+0+0$ phase (broken line).
$\varepsilon_f$ is set to $-0.7$ eV.
}
\label{f08dos}
\end{figure}

Figure \ref{f08dos} shows the density of states (DOS) of the $+$-spin bands near $E_{\rm F}$.
The DOS for the $++00$ phase and that for the $+0+0$ phase are shown.
The  $+$-spin bands  are mainly modified by the magnetic ordering,
and the $-$-spin bands are not modified because $f_\nu$ is not occupied in these phases.
The $p$ bands, which are pushed up by the strong $pf$ mixing,
strongly mixes with the $d$ bands, and the combined effect causes the peak structure
shown in Fig.\ref{f08dos}.
Detailed discussion of the combined effect will be given later.
The peak structure in the $+0+0$ phase appears above $E_{\rm F}$.
Therefore, the combined effect does not contribute to the band energy gain.
In contrast, the peak structure in the $++00$ phase appears
below $E_{\rm F}$  and above $E_{\rm F}$ (the peaks marked by ${\bf a}$ and ${\bf b}$).
This causes the band energy gain.

\begin{figure}[hbtp]
\epsfig{file=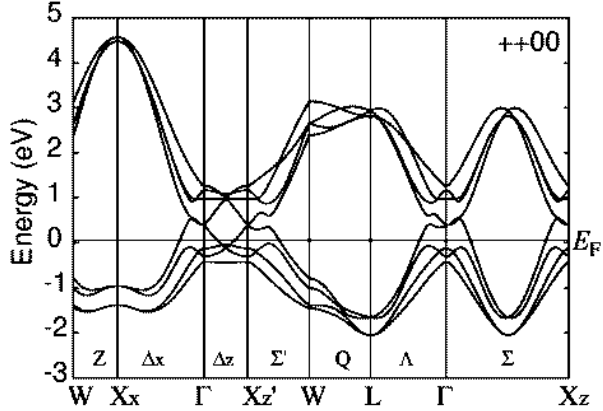,width=8cm}
\caption{
The band structure of the $+$-spin bands for the $++00$ phase.
These bands correspond to the DOS in Fig.\ref{f08dos}.
$X_X$ corresponds to $(1,~0,~0)$ and $X_Z^\prime$ corresponds to $(0,~0,~1/2)$.
The $\Sigma^\prime$ axis connects $(0,~0,~1/2)$ and $(0,~1/2,~1)$.
The other points and axes are the ones of the fcc lattice.
}
\label{f7band}
\end{figure}

Figure \ref{f7band} shows the band structure of the $+$-spin bands for the $++00$ phase.
A part of the $p$ bands is pushed up above $E_{\rm F}$ on the $\Delta_Z$ axis 
due to the $pf$ mixing with the occupied $f$ states.
The $p$ bands have gaps on the $\Delta_Z$ axis caused by the ordering of the $f$ states.
On this axis, the $pf$ mixing effect is maximum,
while the $pd$ mixing disappears due to symmetry.
In contrast, the $pd$ mixing term recovers its value steeply 
when the ${\mib k}$ point moves from the $\Delta_Z$ axis\cite{Kasuya93}.
Since the $p$ bands, which are pushed up by the $pf$ mixing, have 
energy close to the energy of the $d_{xy}$ bands,
the band structure is strongly modified in the vicinity of the $\Delta_Z$ axis.
This fact is seen in Fig.\ref{f7band}. 
On the axes $\Delta_X$ and  $\Sigma^\prime$, 
we can see band splitting effect between the $p$ bands and the $d$ bands.

The $p$ bands, which are pushed up above $E_{\rm F}$, have small band dispersion
along the direction parallel to the $\Delta_Z$ axis.
Then the $pd$ hybridized bands have approximately cylindric equi-energy surfaces
around the $\Delta_Z$ axis.
This enhances the peak structure of DOS due to the van Hove singularity.
We note that  the peak ${\bf a}$ in Fig.\ref{f08dos} corresponds to 
the maximum in Fig.\ref{f7band} just below $E_{\rm F}$ in the $\Delta_X$ axis
(the 3rd band from the low energy side).

\begin{figure}[hbtp]
\epsfig{file=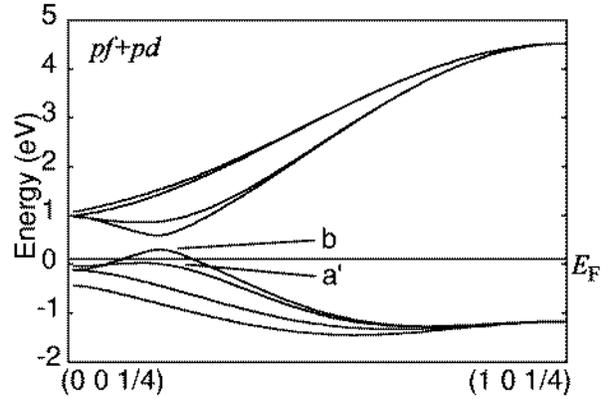,width=8cm}
\caption{
The band structure for the $++00$ phase along 
the $(0,~0,~1/4)-(1,~0,~1/4)$ axis in
the $pf$+$pd$ mixing model.
}
\label{f08f0}
\end{figure}

\begin{figure}[hbtp]
\epsfig{file=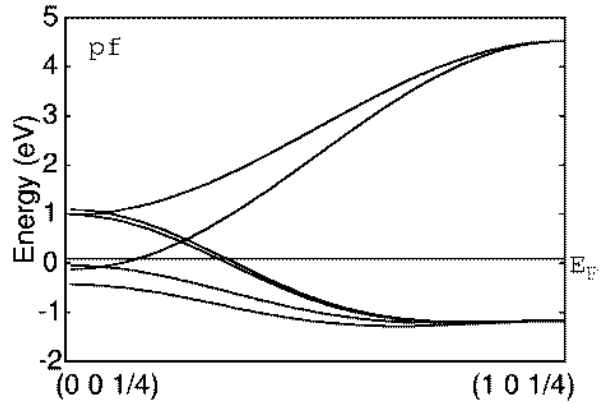,width=8cm}
\caption{
The band structure for the $++00$ phase along 
the $(0,~0,~1/4)-(1,~0,~1/4)$ axis in the $pf$ mixing model.
}
\label{f08f0-}
\end{figure}

Figure \ref{f08f0} shows the band structure for the $++00$ phase 
along the $(0,~0,~1/4)-(1,~0,~1/4)$ axis, 
which is calculated by the $pf$+$pd$ mixing model.
Figure \ref{f08f0-} shows one calculated by the $pf$ mixing model.
The energies of bands in both cases are identical at $(0,~0,~1/4)$, 
because the $pd$ mixing disappears due to symmetry.
When the point moves to the $(1,~0,~1/4)$ direction, 
the band splitting due to the $pd$ mixing is seen in Fig.\ref{f08f0}.
The maximum marked by ${\bf a^\prime}$ in Fig.\ref{f08f0} corresponds to the shoulder 
marked by ${\bf a^\prime}$ in Fig.\ref{f08dos}, 
and the maximum marked by ${\bf b}$ in Fig.\ref{f08f0} corresponds to the 
peak marked by ${\bf b}$ in Fig.\ref{f08dos}.
We note that the point ${\bf a^\prime}$ has maximum type singular point, and 
the point ${\bf b}$ has the saddle type singular point. 

We note that the DOS of the $pf$ mixing model does not have fine structure, 
because of the absence of the $pd$ mixing term.
The gain of the band energy due to the combined effect of the $pf$ mixing and the $pd$ mixing 
around the $\Delta_Z$ axis is important to 
stabilize the double layer structure.

Let us consider the origin of the relative stability between the $++00$ phase and the $+0+0$ phase.
The latter phase has the period characterized by the wave number vector $(0,~0,~1)$.
At the point where the main band gap is formed, 
the $d$ band has energy higher than the Fermi energy of the paramagnetic phase.
Therefore, the main band repulsion due to the combined effect occurs rather higher energy region.
On the other hand, the $++00$ phase is characterized by the wave number vector $(0,~0,~1/2)$.
The gap is formed near the Fermi surface of the $d_{xy}$ bands.
When we remove the reservoir from the present model, 
$E_{\rm F}$ locates just above {\bf b} in Fig.\ref{f08f0},
and the stability of the $+0+0$ phase is relatively enhanced.

\begin{figure}
\epsfig{file=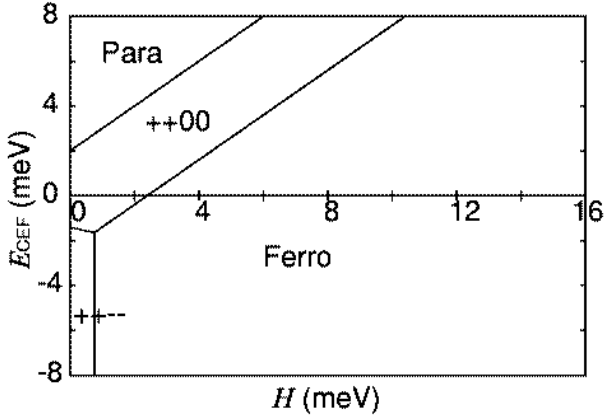,width=8cm}
\caption{
The magnetic phase diagram with the  $f$ level $\varepsilon_f=-0.7$ eV. 
The transverse axis is the field $H$, 
and the ordinate axis is the crystal field splitting $E_{\rm CEF}$ 
between the $\Gamma_7$ layer and the $\Gamma_8$ layer  ($E_{\Gamma_8}-E_{\Gamma_7}$). 
`Para' means the $\Gamma_7$ phase.
The energy of the phases with period-$1$,$2$,$4$ are compared in this figure.
}
\label{f920}
\end{figure}

Figure \ref{f920} shows the magnetic phase diagram under the field 
with the parameter $\varepsilon_f=-0.7$ eV.~
As noted previously, the crystal field splitting 
between the $\Gamma_8$ states and the $\Gamma_7$ states,
\begin{equation}
E_{\rm CEF}=E_{\Gamma_8}-E_{\Gamma_7}
\end{equation}
is a free parameter of the present model.
The energy difference between the ordered states with different fraction of 
the $\Gamma_8$ layer depends on the quantity $E_{\rm CEF}$.
At the same time, the energy gain from the Zeeman interaction is proportional 
to the ferromagnetic moment of the phase.
We have calculated the energy of each phase assuming following equation
\begin{equation}
E=E_{\rm band} + (n_++n_-)E_{\rm CEF} - (n_+ - n_-) H.
\end{equation}
Here, $n_+$ and $n_-$ are the fraction of the $+$ layers and the $-$ layers, respectively.
The $\Gamma_7$ layers are assumed as non-magnetic layers for simplicity.

When the energy of the $\Gamma_8$ layer is enough lower 
than that of the $\Gamma_7$ layer ($E_{\rm CEF} < -1.4$ meV in Fig.\ref{f920}), 
the $++--$ phase becomes the ground state in the $H=0$ case. 
This phase may correspond to the type-IA phase of CeSb.
The phase turns into the ferromagnetic phase in the present calculation, 
when the field is increased. 
The boundary between the $++--$ phase and the ferromagnetic phase 
is about $6$ T in the present calculation, 
and is comparable to the experimental magnitude $4$ T\cite{RossatMignod85}. 
In the calculation of the Zeeman energy, the $g$-value of the 
$\Gamma_8$ states is included as $H \leftarrow g_J \mu_B H$ with $g_J=6/7$.

When the energy of the $\Gamma_7$ layer 
({\it i.e.} the paramagnetic layer in the present model) 
is enough lower 
than that of the $\Gamma_8$ layer ($E_{\rm CEF} > 2$ meV in Fig.\ref{f920}), 
the $\Gamma_7$ phase $(0000)$ becomes the ground state.
This phase turns into the ferromagnetic phase 
by way of the $++00$ phase as the field increases. 

The  phase transition with the parameter $E_{\rm CEF} > 2$ meV 
will mimic that of CeP, 
whose ground state is the antiferromagnetic $\Gamma_7$ phase.
In CeP, the phase successively turns into the states which have 
large fraction of double $\Gamma_8$ layers when the field is increased\cite{KohgiA0}.

\section{Optical Conductivity Spectra}

In this section, we calculate optical conductivity spectra
based on the $pf$+$pd$ mixing model.
The optical conductivity spectra for the $\pm 1$ circular light, $\sigma_\pm(E)$,
of the magnetic ordered states are given by the following expression,

\begin{eqnarray}
\lefteqn{\sigma_\pm(E)}
\nonumber \\
& \propto &
 \int_{\rm BZ} {\rm d}^3 {\mib k}  
\sum_{i, j, n, s, t}
\sum_{
E_i({\mib k}) < E_{\rm F} < E_j({\mib k})
}
\delta(E_j({\mib k})-E_i({\mib k})-E)
\nonumber \\
& \times &
 |<a_j({\mib k})|d_s ({\mib k}+n \Delta {\mib k})> 
 <p_t ({\mib k}+n \Delta {\mib k})|a_i({\mib k})>|^2
\nonumber \\
& \times & w({\mib k} + n \Delta {\mib k}) ~ c^{p \rightarrow d}_\pm(s,t) \nonumber \\
& + & 
 \int_{\rm BZ} {\rm d}^3 {\mib k}
\sum_{i, j, n, s, t} 
\sum_{
E_i({\mib k}) < E_{\rm F} < E_j({\mib k})
}
\delta(E_j({\mib k})-E_i({\mib k})-E)
\nonumber \\
& \times  & 
|<a_j({\mib k})|p_t({\mib k}+n \Delta {\mib k})>
<d_s ({\mib k}+n \Delta {\mib k})|a_i({\mib k})>|^2
\nonumber \\
& \times & w({\mib k} + n \Delta {\mib k}) ~ c^{d \rightarrow p}_\pm(s,t)
,\label{opceq}
\\
\lefteqn{ \Delta {\mib k}  =   (0,~0,~ \frac{2 \pi}{Na}). }
\end{eqnarray}
where $a_j({\mib k})$ stands for the diagonalized band with the band index $j$, 
and BZ means the integration in the Brillouin zone.
$w({\mib k} + n \Delta {\mib k})$ stands for the ${\mib k}$ dependence of the optical transition probability.
The term $w({\mib k} + n \Delta {\mib k})$ is proportional to $(\cos k_x + \cos k_y)^2$, 
since the present $p \leftrightarrow d$ transition is the inter-site process.
However, we assume the term as constant, 
because the saddle type singular points which correspond to the peak structure of the spectra are located
neighboring the $\Delta_Z$ axis where $k_x \simeq 0$ and $k_y \simeq 0$.
$s$ and $t$ represent the indices of the $d$ states ($\alpha$ and $\beta$) and the $p$ states ($\kappa$ and $\nu$) 
respectively.
The terms $c^{p \rightarrow d}_\pm(s,t)$ and $c^{d \rightarrow p}_\pm(s,t)$ stand for 
the selection rule of the optical transition.
They have $1$ for 
$c^{p \rightarrow d}_+(\alpha,\kappa)$, $c^{d \rightarrow p}_+(\beta,\nu)$,
$c^{p \rightarrow d}_-(\beta,\nu)$, $c^{d \rightarrow p}_-(\alpha,\kappa)$,
and $0$ for the other combinations of $s$ and $t$.
We note that the optical transition matrix appears 
only between $p_\kappa$ and $d_\alpha$ states, or between $p_\nu$ and $d_\beta$ states
in the present simplified model.
At the same time, the $+$-spin bands (which consists of $p_\kappa$ and $d_\alpha$) 
do not hybridize with the $-$-spin bands ($p_\nu$ and $d_\beta$).
By the facts, the optical conductivity is separated into the $p \rightarrow d$ components 
and the $d \rightarrow p$ components as expressed in eq.(\ref{opceq}).

\begin{figure}[hbtp]
\epsfig{file=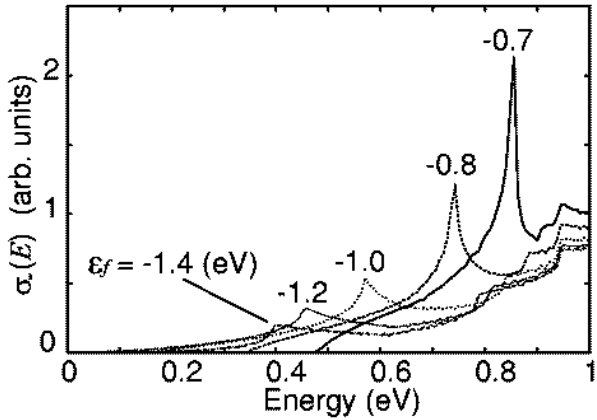,width=8cm}
\caption{
The optical conductivity spectra $\sigma_-(E)$ for the ferromagnetic phase. 
The spectra for $\varepsilon_f=$ $-1.4$, $-1.2$, $-1.0$, $-0.8$, -$0.7$ eV are shown.
The intensity of the $pf$ mixing is proportional to $-1/\varepsilon_f$.
}
\label{f0a040}
\end{figure}

Figure \ref{f0a040} shows the calculated optical conductivity spectra 
for the ferromagnetic phase.
$\sigma_-(E)$ for various $\varepsilon_f$ cases are shown in the figure.
When the $pf$ mixing is large enough, one of the $p$ bands is pushed up above $E_{\rm F}$ around $X_Z$.
Then, the $d \rightarrow p$ transition around $X_Z$ comes to be possible, 
and the transition contributes to $\sigma_-(E)$.
The band in the ferromagnetic phase has a saddle type singular point 
of the joint density of states at $X_Z$, 
and the peak structure, which corresponds to the transition around this point, appears in $\sigma_-(E)$.
On the other hand, 
$\sigma_+(E)$ is essentially equivalent to that of the paramagnetic phase
consistently with the experimental results by Kimura {\it et al.}\cite{Kimura00,Kimura00PB}.

The energy of the peak 
$E_{\rm peak}$ in Fig.\ref{f0a040} becomes higher as the strength of the $pf$ mixing effect, 
which is proportional to $-1/\varepsilon_f$, becomes larger.
At the same time, the strength of the peak increases as the intensity 
of the $pf$ mixing effect increases, 
because the ${\mib k}$-space which contributes to the peak 
becomes wider. 

\begin{figure}
\epsfig{file=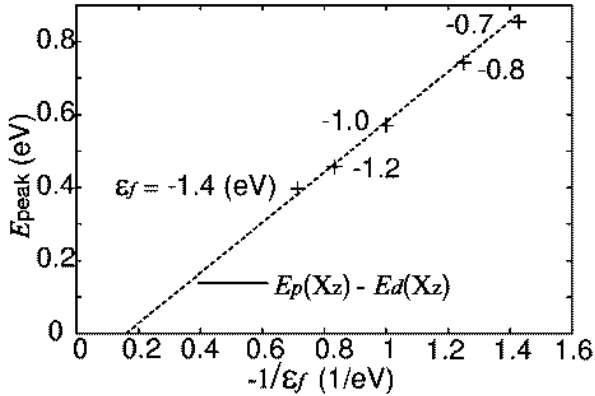,width=8cm}
\caption{
The energy of the peak $E_{\rm peak}$ of the ferromagnetic phase 
versus the strength of the $pf$ mixing effect ($-1/\varepsilon_f$).
The peak energy is plotted by `$+$' sign. 
The broken line represents the energy difference  at  $X_Z$
between the $p$ band $E_p$ and the $d$ band  $E_d$. 
}
\label{fa50}
\end{figure}

Figure \ref{fa50} shows the relation between the energy of the peak $E_{\rm peak}$
of the ferromagnetic phase and  $-1/\varepsilon_f$.
The marks `$+$' in the figure are roughly on the line of $E_p(X_Z)-E_d(X_Z)$, 
which is the energy difference between the $p$ band and the $d$ band at $X_Z$. 

Using the relation between the energy of the peak and $-1/\varepsilon_f$,
we can determine  $-1/\varepsilon_f$
from the experimental results\cite{Kimura00}.
We obtain the value of the $f$ level as $\varepsilon_f=-1.35$ eV 
from Fig.\ref{fa50}, where the peak appears at $0.4$ eV.
In later calculation of the optical conductivity,  we employ this value.
In this case, 
the energy difference in Fig.\ref{f940}
becomes about half of 
that of the 
$\varepsilon_f=-0.7$ eV
case,
and this case also holds
the condition of the relative stability of the double layer structure.
The value of $\varepsilon_{f}$
depends on the magnitude of the $pf$ hybridization matrix element\cite{Sakai84}.
On the other hand, the energy of the peak directly corresponds to the band energy
difference $E_p(X_Z)-E_d(X_Z)$.
Therefore, the present choice means that 
the later calculation is carried out based on the band parameters
which give the band energy 
difference $0.4$ eV in the ferromagnetic phase at $X_Z$.
When we use this readjustment, the calculated
spectral change due to the change of the magnetic ordering
will scarcely depend on the detailed choice of the band parameters.

We note that the transition around $X_Z$ becomes a maximum type singular point
in the $pf$ mixing model.
Therefore, we do not have peak structure at this energy,
if the $pd$ mixing is neglected.
We also note that the reservoir bands are not modified by magnetic ordering, 
and their optical conductivity spectra are equivalent to those of the paramagnetic phase.
The low energy peak structure
which appears in the ordered states will be scarcely affected 
by the contribution from the transitions among the reservoir bands.

\begin{figure}
\epsfig{file=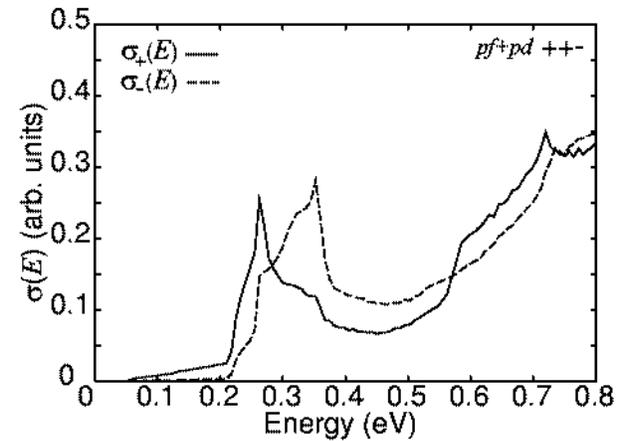,width=8cm}
\caption{
The optical conductivity spectra for the $++-$ phase. 
}
\label{fan0}
\end{figure}

Figure \ref{fan0} shows the calculated optical conductivity spectra for the $++-$ phase
with  $\varepsilon_f=-1.35$ eV.~
Both $\sigma_+(E)$ and $\sigma_-(E)$ have a peak and a shoulder structure 
which are located at  $0.26$ eV and at $0.37$ eV. 
The peaks mainly consist of $d \rightarrow p$ transition, 
and the shoulders mainly consist of $p \rightarrow d$ transition.
Both the peaks and the shoulders have the lower energy than the energy of 
the peak of the ferromagnetic phase. 
These behaviors are consistent with the observation\cite{KimuraPC}.

In the $++-$ phase, the $pf$ mixing pushes up both the $+$-spin bands and the $-$-spin bands, 
and the $pd$ band repulsion makes saddle type singular points which correspond to the 
structure in Fig.\ref{fan0}.
In contrast to the peak structure of the ferromagnetic phase, 
the transition around $X_Z$ does not correspond to the structure.

\begin{figure}
\epsfig{file=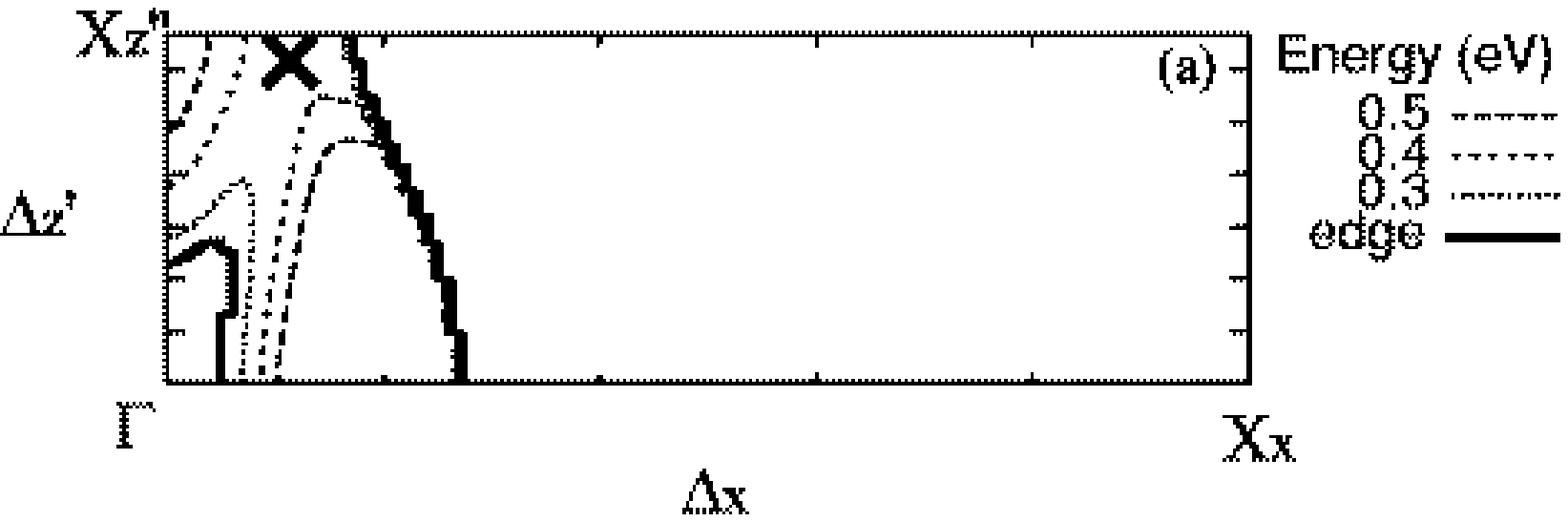,width=8cm}
\\
\epsfig{file=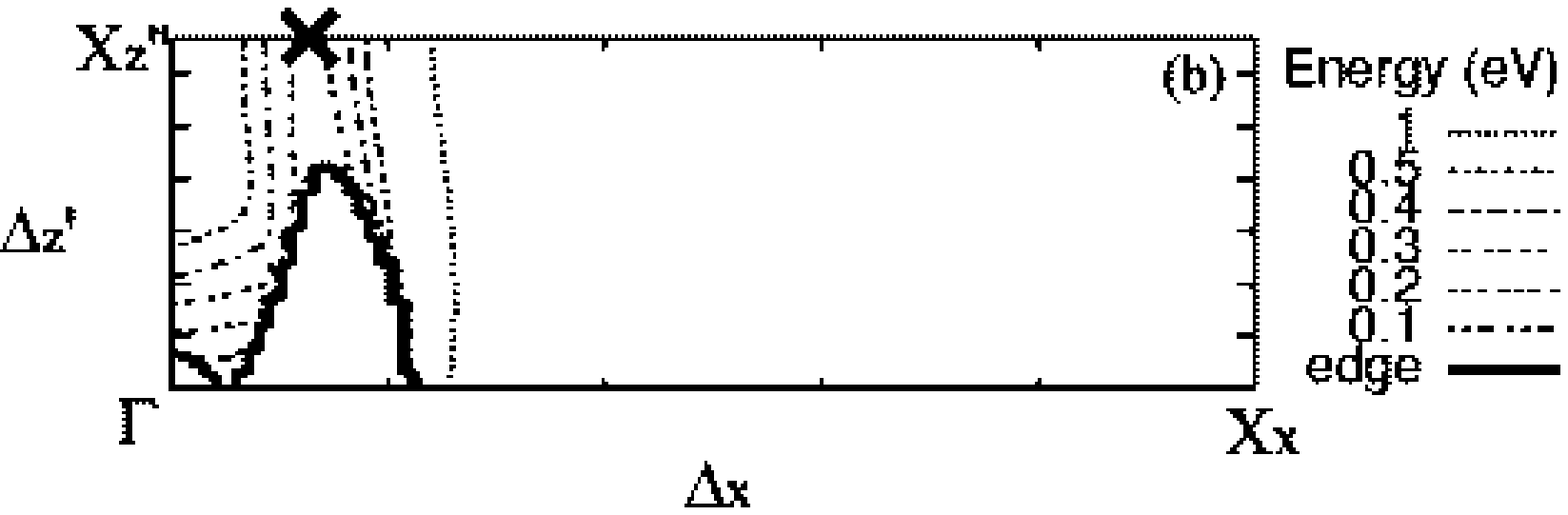,width=8cm}
\caption{
The equi-energy contour map for the optical transition in  the $++-$ phase.
The transition of the $+$-spin bands shown in (a) contributes mainly to the peak at $0.26$ eV of $\sigma_{+}(E)$ 
and 
the transition of the $-$-spin bands shown in (b) contributes mainly to the peak at $0.37$ eV of $\sigma_{-}(E)$.
Both peaks correspond to the transition from maxima just below $E_{\rm F}$ to minima just above $E_{\rm F}$
near $X$ on the $\Sigma^\prime$ axis in Fig.10 of ref.\citen{KimuraPC}.
The equi-energy contours of $E=$ $0.3$, $0.4$, $0.5$ eV (a) 
and those of $E=$ $0.1$, $0.2$, $0.3$, $0.4$, $0.5$, $1.0$ eV (b) are plotted.
The region where the optical transition does not occur is 
closed by the contour named `edge'.
$X_Z^{\prime\prime}$ is $(0,~0,~1/3)$, and the $\Delta_Z^\prime$ axis connects $\Gamma$ and $X_Z^{\prime\prime}$.
%
The saddle type singular points are marked by `$\times$' in the figures.
}
\label{far0as0}
\end{figure}

Figure \ref{far0as0} shows the contour map of the interband transition energy for the $++-$ 
phase in the ${\mib k}$-space.
This map has approximately  axial symmetry around the $\Delta_Z^\prime$ axis.
The peak of $\sigma_+(E)$ and the shoulder of $\sigma_-(E)$ 
at $0.26$ eV in Fig.\ref{fan0} correspond to the saddle type singular point
shown by `$\times$' in Fig.\ref{far0as0}(a).
The former corresponds to the $d \rightarrow p$ transition around the point, and the 
latter corresponds to the $p \rightarrow d$ transition around the same point.
The peak of $\sigma_-(E)$ at $0.37$ eV in Fig.\ref{fan0} corresponds to the 
$d \rightarrow p$ transition around the saddle type singular point shown by `$\times$' 
in Fig.\ref{far0as0}(b), 
and the shoulder of $\sigma_+(E)$ corresponds to the 
$p \rightarrow d$ transition around the same point.

\begin{figure}[hbtp]
\epsfig{file=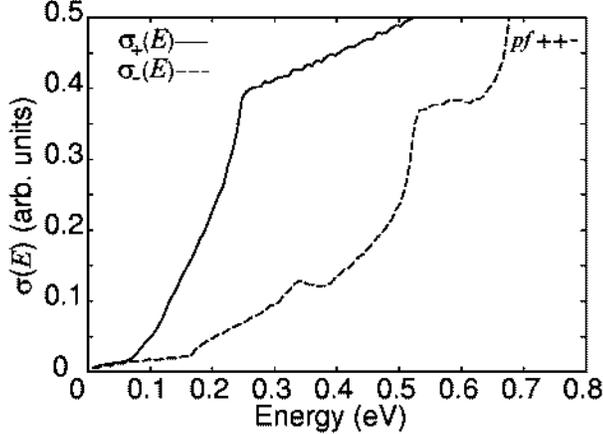,width=8cm}
\caption{
The optical conductivity spectra for the $++-$ phase of the $pf$ mixing model.
}
\label{faq0}
\end{figure}

In Fig.\ref{faq0}, we show
the optical conductivity spectra for the $++-$ phase calculated by neglecting the $pd$ mixing 
term (i.e. the $pf$ mixing model).
There appears low energy transitions, caused partly by the Brillouin zone folding effect.
This case does not have peak structure.
That is, the peaks of the optical conductivity spectra originate 
from the saddle type singular points caused by the combined effect 
of the $pf$ mixing and the $pd$ mixing effect.

\begin{figure}[hbtp]
\epsfig{file=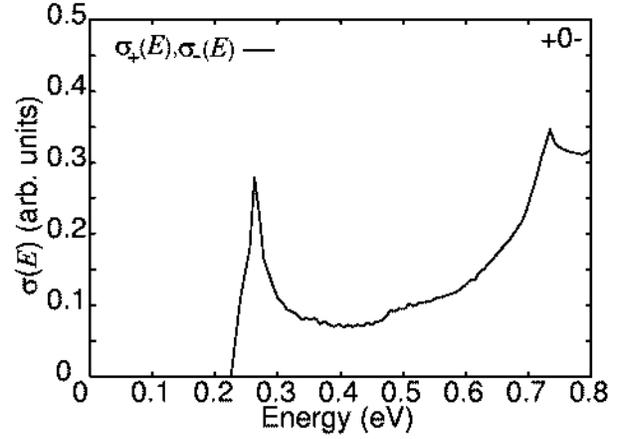,width=8cm}
\caption{
The optical conductivity spectra for the $+0-$ phase.
$\sigma_+(E)$ and $\sigma_-(E)$ have the same spectrum.
}
\label{f0am0}
\end{figure}

Figure \ref{f0am0} shows the optical conductivity spectra for the $+0-$ phase 
with $\varepsilon_f=-1.35$ eV.~
$\sigma_+(E)$ and $\sigma_-(E)$  are identical, 
and they have a peak at $0.26$ eV.
The peak structure in the $+0-$ phase and
the structure of $\sigma_+(E)$ in the $++-$ phase shown in Fig.\ref{fan0} have the same energy.
It is because the 
$-$-spin bands in the $+0-$ phase are identical to those 
in the $++-$ phase.
The difference is on
the centroid of the peak, which is shifted to higher energy side in
the $++-$ phase than that of the $+0-$ phase
due to the contribution from the $+$-spin bands.
The experimental result of the spectra for this phase has not been reported.
We expect that the features shown in the figure will be observed.

\begin{figure}[hbtp]
\epsfig{file=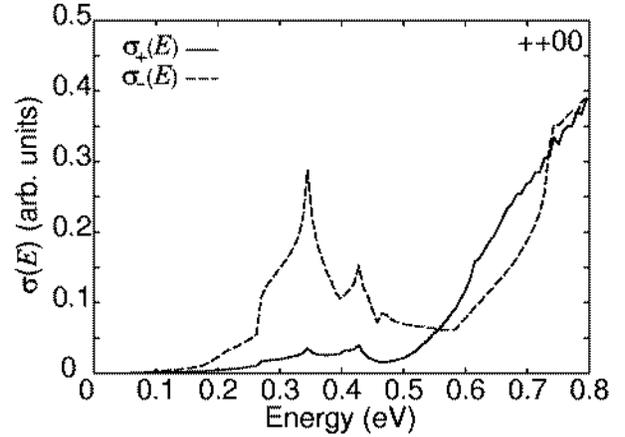,width=8cm}
\caption{
The optical conductivity spectra for the $++00$ phase.
}
\label{f0au0}
\end{figure}

Figure \ref{f0au0} shows the optical conductivity spectra for the $++00$ phase
with $\varepsilon_f=-1.35$ eV.
There appears a double peak structure, and
both the peaks at $0.35$ eV and at $0.43$ eV in $\sigma_-(E)$
consist of the $d \rightarrow p$ transition.
$\sigma_+(E)$ has the structure corresponding to that in $\sigma_-(E)$
with weak intensity.
These results are consistent with the observation\cite{KimuraPC}. 

\begin{figure}[hbtp]
\epsfig{file=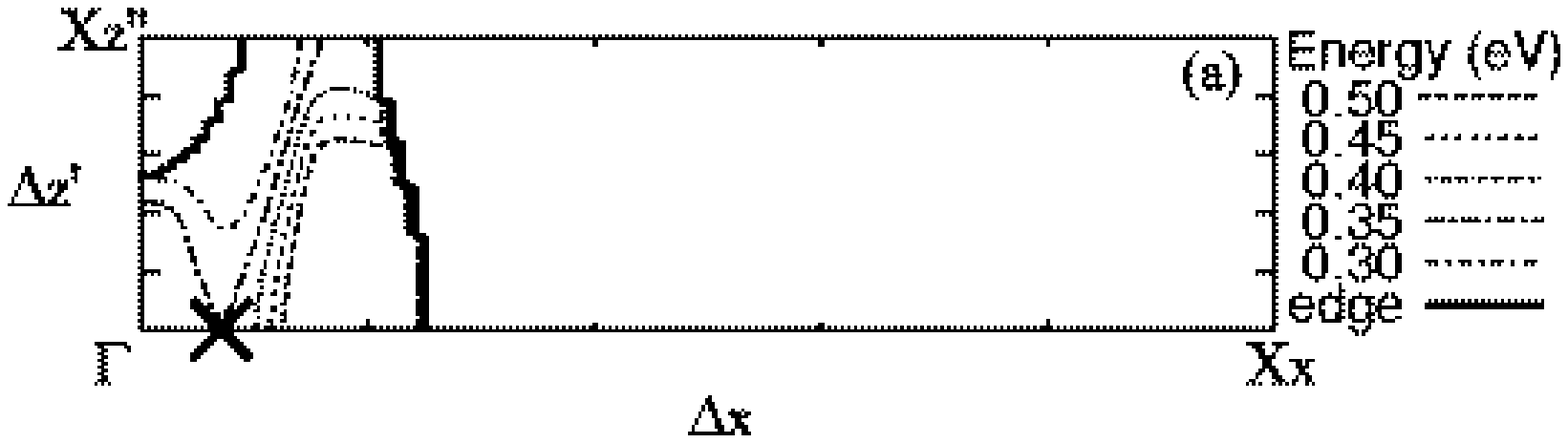,width=8cm}
\\
\epsfig{file=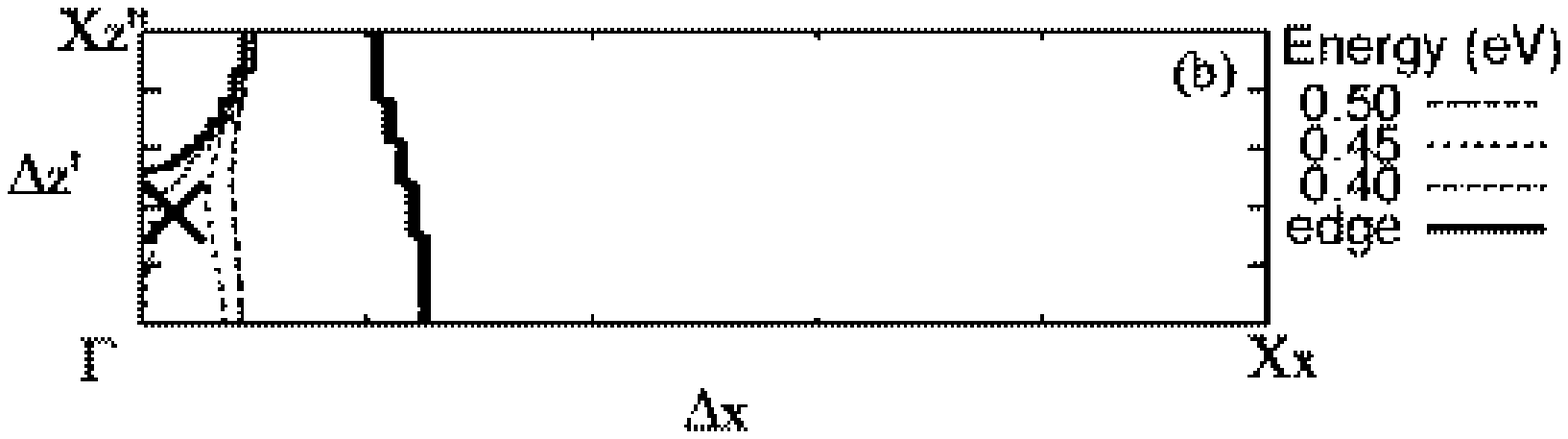,width=8cm}
\caption{
The equi-energy contour map for the optical transition in  the $++00$ phase.
Upper figure (a), for which there appears the peak at $0.35$ eV, 
corresponds to the transition from the third lowest band to 
the fourth lowest band shown in Fig.\ref{f7band},
and the lower figure (b), for which there appears the peak at $0.43$ eV, 
corresponds to that from the second lowest band to the 
fourth lowest band shown in Fig.\ref{f7band}.
$X_Z^{\prime\prime}$ is $(0,~0,~1/4)$, and the $\Delta_Z^\prime$ axis connects
$\Gamma$ and $X_Z^{\prime\prime}$. 
The saddle type singular points are marked by `$\times$' in the figures.
The equi-energy contours of $E=$ $0.3$, $0.35$, $0.4$, $0.45$, $0.5$ eV (a) 
and those of $E=$ $0.4$, $0.45$, $0.5$ eV (b) are plotted.
The region where the optical transition does not occur is 
closed by the contour named `edge'.
}
\label{f0aw0}
\end{figure}

In Fig.\ref{f0aw0}, we indicate 
the saddle type singular points neighboring the $\Delta_Z^\prime$ axis 
which correspond to the peaks of $\sigma_-(E)$ shown in Fig.\ref{f0au0}.
Similarly to the interband transition structure of the $++-$ phase, 
the  dispersion of the transition energy is small 
in the direction parallel to the $\Delta_Z^\prime$ axis 
around the saddle type singular points, 
and the cylindrical equi-energy region 
appears
in the $++00$ phase.

\section{Summary and Discussion}

We have demonstrated that the combined effect of the $pf$ mixing and the $pd$ mixing
is important to stabilize the double layer structure 
which is widely observed in the magnetic phase diagram of CeSb.
A part of the $p$ bands is pushed up along the $\Delta_Z$ axis 
by the $pf$ mixing with occupied $f$ state.
These $p$ bands strongly mix with $5d$ bands around the $\Delta_Z$ axis 
through the $pd$ hybridization.
This causes several bands which have saddle type singular points neighboring the axis.
The energy gain due to the $pf$ mixing is substantiated by the $pd$ hybridized bands.
The same bands give the saddle type singular points of the joint density of states
of the optical transition in the magnetic ordered states.
Characteristics of the magneto-optical spectra for various phases of CeSb are 
reproduced by the calculation.

The energy of the peak structure of the optical conductivity in the ferromagnetic state 
corresponds to the excitation energy between the pushed up $p$ state and the $d$ state 
at $X_Z$.
Therefore the peak structure in the ferromagnetic phase of La$_x$Ce$_{1-x}$Sb 
is expected to shift to low energy side as $x$ becomes large, 
because the $pf$ mixing effect decreases.
The peak intensity will also decrease.
The situation similar to that shown in Fig.\ref{f0a040} will be shown.

In the present paper, we have employed the simplified model 
to carry out numerical calculation with sufficient accuracy for the band energy 
of various magnetic phases.
The same simplified model accounts the characteristics of the magneto-optical spectra of
each magnetic phase.
However, the bands treated as the reservoir will also contribute to the spectra
in various way. 
For example, the transition between the reservoir and the states
which are modified strongly by the magnetic ordering will be also affected.
Such transition terms are not considered in the present model.
Further studies including total bands are necessary in quantitative analysis of the magneto-optical spectra.

Recently, Iwasa {\it et al.} have proven by the fine X-ray diffraction experiment 
that the `paramagnetic layer' consists of the $\Gamma_7$ states\cite{Iwasa00}.
The present model ascribing the paramagnetic layer as the $\Gamma_7$ states will be justified.
We have treated the layer as the inert state in the band calculation.
In the strict sense, the $\Gamma_7$ states have hybridization term with the band states.
However, it is not large for the states near the Fermi energy.
The essential features of the present result will not be modified even when we include
the hybridization term of the $\Gamma_7$ states.
We expect that the Kondo effect for the paramagnetic state\cite{Martinez} will not be so important.

We have shown the stability of the double layer structure on the basis of the band calculation assuming 
the long range periodicity.
The states which include the double layer structure seems to be more stable 
than the states with isolated layers even when we consider the periodicity of eight layers\cite{ORBITAL}.
Therefore, the explanation of the stability of the double layer structure 
on the basis of the local electronic structure, in some sense, is desirable. 
In the experiments on CeP, we observe phases which have very long periodicity 
with a small number of double layer in the sea of the paramagnetic state\cite{KohgiA0}.
Such a study on the isolated double layer is retained in the future.
Nevertheless, we note that, in the short periodicity cases, 
the present band calculation picture will be applicable.

\section*{Acknowledgement}
The authors would like to thank 
K. Ueda for encouraging discussions, 
S. Kimura for comments on experiments,
Y. Kuramoto for valuable comments, and 
H. Harima, Y. Shimizu and Y. Kaneta 
for supporting numerical calculation.
Numerical calculation was partly performed in the 
Computer Center of Tohoku University,
the Supercomputer Center of Institute for Solid State Physics (University of Tokyo)
and
The Institute of Scientific and Industrial Research (Osaka University).


\end{document}